# Fast High-Fidelity Measurements of the Ground and Excited States of a dc SQUID Phase Qubit

T. A. Palomaki, S. K. Dutta, R. M. Lewis, Hanhee Paik, K. Mitra, B. K. Cooper, A. J. Przybysz,

A. J. Dragt, J. R. Anderson, C. J. Lobb, and F. C. Wellstood

Center for Superconductivity Research

and

Joint Quantum Institute

Department of Physics, University of Maryland, College Park, Maryland 20742-4111

## Abstract

We have investigated the fidelity and speed of single-shot current-pulse measurements of the three lowest energy states of the dc SQUID phase qubit. We apply a short ($2ns$) current pulse to one junction of a Nb/AlO$_x$/Nb SQUID that is in the zero voltage state at 25 mK and measure if the system switches to the finite voltage state. By plotting the switching rate versus pulse size we can determine average occupancy of the levels down to 0.01%, quantify small levels of leakage, and find the optimum pulse condition for single-shot measurements. Our best error rate is 3% with a measurement fidelity of 94%. By monitoring the escape rate during the pulse, the pulse current in the junction can be found to better than 10 nA on a 0.1 ns time scale. Theoretical analysis of the system reveals switching curves that are in good agreement with the data, as well as predictions that the ultimate single-shot error rate for this technique can reach 0.4% and the fidelity 99.2%.

## I. Introduction

Recent advances in Josephson-junction qubits have gone a long way towards establishing the possibility of using superconducting devices for quantum computation. Some key demonstrations include the observation of Rabi oscillations [1-6], successful coupling of two or more qubits [7-10], improved coherence times [11], and state tomography [12]. Nevertheless, many necessary conditions remain unsatisfied, including the need for a readout technique that is fast enough, and has a low enough error rate that it could be used to implement error correction in a logical qubit. Several techniques for performing state readout have been introduced for Josephson junction qubits, including tunneling [13], resonant microwave pumping to an auxiliary level [2], pulsed current [14-15], the Josephson bifurcation amplifier [16-17], and resonant reflectometry [18]. Here we report experimental results and theoretical analysis of the speed and ultimate fidelity of the pulsed current readout scheme when used to measure a dc SQUID phase qubit. We also demonstrate how the technique can be used to measure undesired leakage to higher energy levels and characterize the current profile during the pulse.

## II. Pulsed readout of dc SQUID qubits

Figure 1 shows a schematic and photograph of our dc SQUID phase qubit. With the proper choice of circuit parameters [19] the qubit junction (junction 1) acts effectively as a single well-isolated, current-biased Josephson junction phase qubit [20]. The qubit junction is placed in series with a relatively large inductor $L_1$, and this combination is placed in parallel with a small inductor $L_2$ and an isolation junction $J_2$. As first pointed out by Martinis *et al*. [2], this configuration forms a current divider yielding broadband isolation of the qubit junction from current noise on the bias leads. For good isolation, we choose $L_1 >> L_2$, producing a dc SQUID with a large inductive asymmetry. By simultaneously ramping the bias current $I_b$ and flux $\Phi_a$ applied to the loop in the proper proportion, the current through each junction can be independently controlled. As in the Josephson junction phase qubit, the two lowest energy levels ($n = 0$ and 1) in one well of the potential form the computational basis.

Following other groups [14-15], we perform state readout by applying a brief current pulse to the qubit junction and checking if the device escapes to the voltage state: Switching of the qubit junction causes the isolation junction to switch, leading to a steady voltage of about 2.8 mV (Nb/AlO$_x$/Nb junction) across the bias leads. The technique exploits the different escape rates of the levels; each successive energy level has a tunneling rate that is two to three orders of magnitude faster than the level below it (see Fig. 2). As Silvestrini *et al.* first showed experimentally, different energy levels become unstable, or reach a specified high escape rate, at different current levels through the junction [21]. Thus by carefully choosing the size of the current pulse, one can ensure with a high probability that levels n and higher will tunnel, while $n-1$ and below will not.

Occupancy of $n = 2$ and higher levels is undesirable unless the levels are being used as auxiliary states for gate manipulation or readout [2,22]. Unfortunately, it is relatively easy in this system to get leakage to higher levels. Since the levels are anharmonic, although not strongly so, pumping between specific states can be achieved by applying resonant microwaves to induce Rabi oscillations. High speed operation is desirable, as there is less chance for the system to decohere, and this generally requires the use of higher microwave power. However, high power causes resonances to broaden, leading to a potentially significant population in higher levels. The effects of higher levels are readily seen in Rabi oscillations as a continuous increase in the escape rate with power, up to and beyond the escape rate from 100% occupancy of $n = 1$ [23]. To quantify these undesirable leakage effects, it is essential for the state measurement technique to be able to measure small populations in the upper level $n = 2$, as well as the qubit levels $n = 0$ and $n = 1$.

## II. Optimizing the state detection process

To analyze the state detection process, we begin by introducing conditional probabilities $P_{ij}$, where $i$ corresponds to the initial quantum state of the system, and $j$ indicates whether the system has switched into the voltage state ($j = 1$ indicates that the system has switched, $j = 0$ indicates that it has remained in its initial quantum state.). When the system is in state $n = 1$, $P_{11}$ is the probability that a correct detection occurs, i.e. $P_{11}$ is the probability that the state $n = 1$ switches to the voltage state during the current pulse. Similarly, we will define $P_{00}$ as the probability of correctly reporting a 0 when the system is in the state $n = 0$, i.e. the state $n = 0$ does <u>not</u> switch to the voltage state during a current pulse. Four main types of errors will occur:

(i) False detection of the state $n = 1$ when the system was in the state $n = 0$, corresponding to switching to the voltage state when the system was in $n = 0$, occurs with probability $P_{01}$.

(ii) Missed detection of the state $n = 1$, corresponding to not switching when the system is in the state $n = 1$, occurs with probability $P_{10}$.

(iii) False detection of the state $n = 1$ when the system was in the state $n = 2$, corresponding to switching from $n = 2$ to the voltage state, occurs with probability $P_{21}$.

(iv) False detection of the state $n = 0$ when the system is in the state

$n = 2$, corresponding to not switching when in the state $n = 2$, occurs with probability $P_{20}$.

Additional errors will occur if there is occupancy in higher levels, *i.e.* $n = 3$ or $4$. However, this is a significant effect only at very high microwave drive power, which we ignore here. We note that conservation of probability forces some constraints, in particular

$$1 = P_{01} + P_{00} = P_{11} + P_{10} = P_{21} + P_{20} \quad . \qquad [1]$$

It is also important to realize that in general $P_{01} \neq P_{10}$ and that under typical pulse conditions $P_{21} \gg P_{20}$.

Optimizing the detection process involves maximizing the probability of detecting the anticipated signal while minimizing the errors. We can optimize the process by varying the duration and height of the current pulse, since this affects the probabilities $P_{ij}$ differently.

In order to optimize the pulse applied to the qubit, we need to make some assumptions about the initial probabilities $P_i$ of being in the $i^{th}$ quantum state. For simplicity, we will assume that the qubit is equally likely to be in $n = 0$ or $n = 1$ and that there is no probability to be in $n = 2$, *i.e.* $P_0 = P_1 = 0.5$ and $P_2 = 0$. This assumption corresponds to the situation where there is no *a priori* information about the state of the qubit before the measurement. With this choice, the probability of correctly detecting the state of the system is

$$S = P_0 P_{00} + P_1 P_{11} = (P_{00} + P_{11})/2 \quad . \qquad [2]$$

We can think of $S$ as being the average signal. The corresponding probability of making an error is

$$N = P_0 P_{01} + P_1 P_{10} = (P_{01} + P_{10})/2, \qquad [3]$$

and we can consider $N$ as being the average noise.

Using Eqs. [1-3], we can write the signal-to-noise ratio as

$$S/N = \frac{(2 - P_{01} - P_{10})}{P_{01} + P_{10}} = \frac{2}{P_{01} + P_{10}} - 1 = \frac{1}{N} - 1 \qquad [4]$$

and define the measurement fidelity $F$ as [24]:

$$F \equiv 1 - P_{01} - P_{10} = 1 - 2N \qquad [5]$$

While the fidelity mainly has been used in discussions of Rabi oscillations, we believe the closely related concept of the average error $N$ is a more useful concept for error correction.

Examination of Eqs. [3-5] reveals that minimizing $N$ will simultaneously maximize the $S/N$ ratio and $F$. Since $P_{01}$ and $P_{10}$ can be obtained from measurements on a device as a function of pulse parameters at any bias point, Eqs. [3-5] can be used to find the optimum detection conditions from experimental data. To understand the ultimate limits to the pulse detection technique, we first analyze the system analytically using approximate forms for $P_{01}$ and $P_{10}$ and then compare the results to full numerical simulations and our measurements.

To proceed, we express the error probabilities in terms of the escape rates. Using the WKB approximation and a cubic approximation to the potential, the escape rate of the $n^{th}$ level is [25]

$$\Gamma_n = \frac{\omega_p}{n!\sqrt{2\pi}} (432 N_s)^{n+1/2} \exp(-7.2 N_s) \qquad [6]$$

where

$$N_s \cong \frac{2^{3/4}}{3} \left(\frac{E_J}{E_c}\right)^{1/2} \left(1 - \frac{I}{I_o}\right)^{5/4} \qquad [7]$$

is the approximate number of energy levels with energy less than the well's barrier height, $I_0$ is the critical current of the qubit junction, $I$ is the current flowing through the junction, $E_j = I_0 \Phi_0 / 2\pi$ is the Josephson coupling energy and $E_c = e^2/2C_1$ is the electrostatic charging energy of the qubit junction's capacitance $C_1$.

From the escape rates, we can determine the probability that the system has escaped to the voltage state. To proceed, we adopt a simple model of the pulse: We assume that the current pulse starts at $I_a$ where there is negligible probability of escape, and then at $t = 0$ ramps to a value $I(t) = I_a + I_p$ where it holds steady, and then returns to $I_a$ after a time $\tau$ has passed. If the current does not vary too rapidly during the pulse rise and fall, then it will act adiabatically [25,26] on the system and the only changes in the occupancy of the levels will be due to tunneling. The average escape rate at any time during the pulse will then be well-represented by

$$<\Gamma> = P_0(t)\Gamma_0(I_1(t)) + P_1(t)\Gamma_1(I_1(t)) + P_2(t)\Gamma_2(I_1(t)) \qquad [8]$$

where $P_0$, $P_1$, and $P_2$ are the populations of the $n = 0$, $n = 1$, and $n = 2$ state, respectively.

If we ignore the rise time and assume the pulse acts adiabatically, and that the system starts with $P_0 = 1$ at $t = 0$, the cumulative probability $P_{01}(\tau)$ that the state $n = 0$ has escaped (detected as a 1) after the time $\tau$ has passed is

$$P_{01}(\tau) = 1 - \exp\left(-\int_0^\tau \Gamma_0(t')dt'\right) = 1 - \exp(-\Gamma_0 \tau) = 1 - y \qquad [9]$$

where $y = \exp(-\Gamma_0 \tau)$ and $\Gamma_0$ is the escape rate from the $n = 0$ state during the pulse. Similarly, the cumulative probability $P_{11}(\tau)$ that the state $n = 1$ has escaped (detected as a 1) after the time $\tau$ has passed is

$$P_{11}(\tau) = 1 - \exp\left(-\int_0^\tau \Gamma_1(t')dt'\right) = 1 - \exp(-\Gamma_1 \tau) = 1 - \exp(-u\Gamma_0 \tau) = 1 - y^u \qquad [10]$$

where $\Gamma_1$ is the escape rate from the first excited state during the pulse, and we have defined

$u = \Gamma_1/\Gamma_0$ as the ratio of the escape rates during the pulse. From Eqs. [1] and [10], we find

$$P_{10}(\tau) = 1 - P_{11}(\tau) = y^u \quad . \tag{11}$$

Equations [3]-[5] can then be written in the form

$$N = (1 - y + y^u)/2 \quad , \tag{12}$$

$$S/N = \frac{2}{(1 - y + y^u)} - 1 \quad , \tag{13}$$

$$F = y - y^u \quad . \tag{14}$$

Minimizing $N$ with respect to $y$, and thus implicitly with respect to the size of the current pulse or the pulse time $\tau$, one finds that the minimum error $N$ occurs for

$$y^{opt} = (1/u)^{1/(u-1)} \quad . \tag{15}$$

This implies that for the optimum pulse height

$$\Gamma_1 \tau = u \ln(u)/(u-1) \approx \ln(u) \tag{16}$$

since $\Gamma_1 \gg \Gamma_0$. The optimum values for the error rate $N$, the $S/N$ ratio, and fidelity then become

$$N^{opt} = \frac{1}{2}\left(1 - \left(\frac{1}{u}\right)^{\frac{1}{u-1}} + \left(\frac{1}{u}\right)^{\frac{u}{u-1}}\right) \quad , \tag{17}$$

$$(S/N)^{opt} = \frac{2}{1 - \left(\frac{1}{u}\right)^{\frac{1}{u-1}}\left(\frac{u-1}{u}\right)} - 1 \quad , \tag{18}$$

$$F^{opt} = \left(\frac{1}{u}\right)^{\frac{1}{u-1}} - \left(\frac{1}{u}\right)^{\frac{u}{u-1}} \quad . \tag{19}$$

Examination of Eqs. [16]-[19] reveals that the optimum values only depend on $u = \Gamma_1/\Gamma_0$ during the pulse. Figure 3 shows plots of $N$ and $S/N$ for typical values of $u$. We find that $\log (S/N)^{opt}$ varies nearly linearly with $\log u$ while $N^{opt}$ varies approximately inversely with $u$. Since the maximum $u$ is about 1000 for typical junction parameters, we conclude from Fig. 3(b) that the ultimate minimum probability of making a single-shot detection error is $N \cong 0.004$ with this technique. This level of measurement error is much better than reported in any superconducting device so far and is low enough to be compatible with some error correction schemes [27]. However, actual devices have tended to have considerably lower values of $u$ because the ratio $u = \Gamma_1/\Gamma_0$ decreases at the large current values needed to achieve the high escape rates required for short measurement times (see Fig. 2).

We also note that at the optimum bias point, the error probabilities, $P_{10}$ and $P_{01}$, are not the same (see Table 1). For $u = 250$ with the optimum pulse, we find $N = 0.013$ and the $F = 0.974$. In this case the dominant error is $P_{01} = 0.022$ while $P_{10} = 0.0035$ is about 7 times smaller. For comparison, if $u = 750$, the average error falls to $N = 0.005$, the dominant error is $P_{01} = 0.009$ while $P_{10} = 0.0015$ is six times smaller, and $F = 0.99$.

### III. Measurements of the state of a dc SQUID phase qubit

To verify our understanding of the detection process, we tested a dc SQUID phase qubit [see Fig. 1(a)] made by Hypres, Inc., [28] from a Nb-AlO$_x$-Nb trilayer with their $30 A/cm^2$ process. The qubit junction had a nominal area of $(10 \mu m)^2$, a critical current $I_0 = 19 \mu A$ and capacitance $C_1 = 4.3 pF$. For measurements, the SQUID was mounted in an Al box that was connected to the mixing chamber of an Oxford Instruments model 200 dilution refrigerator with a base temperature of about 25 mK. All lines, except the pulse/microwave line, were filtered using custom rf and microwave Cu powder filters [29].

The SQUID was first characterized electrically by measuring the switching current as a function of the applied flux. The device had multiple flux states and flux shaking [19] was used to initialize the flux before each switching measurement. For measurements on the qubit junction, a simultaneous current and flux bias were applied so that current was driven through the qubit junction and not the isolation junction.

Next, pulses were applied to the qubit through a 2 fF on-chip capacitor [$c_\mu$ in Fig 1(b)] that was also used for applying microwave power to excite transitions. Some of the pulse current flows through the isolation junction, but very little, since $L_1$ is much larger than the Josephson inductance of the qubit junction. To form a short pulse ($< 2ns$) at the qubit, we used the capacitor to differentiate the rising edge of a slower current pulse. We monitor the output voltage across the SQUID using a low-noise amplifier, keep track of whether or not the device switches, and resolve the timing of the individual escape events to better than 100 ps. Repeating this for many switching events, we can construct the average probability of switching for a given pulse size and the escape rate as a function of time during the pulse.

The pulsed method presents one difficulty: The current in the qubit during the pulse differs greatly from the current applied at the top of the refrigerator. What is needed is a way to determine the actual pulse current in the qubit vs. time. To calibrate the pulses, we make use of the escape rate of the system. Figure 2 shows the measured escape rate, $<\Gamma>$, of the qubit junction when cooled to 25 mK. Also plotted are fitted results (dotted lines) for $\Gamma_0$, $\Gamma_1$, $\Gamma_2$ using numerical analysis comparable to Eq. [6], for an effective qubit critical current, $I_0 = 17.957 \mu A$. Based on the device parameters, we expect the single junction cubic approximation model to provide a good description of the escape rates [19]. Quantum simulations have also verified this approximation is valid [30]. The disagreement between the measured $<\Gamma>$ and expected $\Gamma_0$ is due to high-frequency noise producing a very small population in $n=2$. Since the escape rate of the second excited state is approximately $\Gamma_2 = 10^8 /s$ in the region of the disagreement, even a small $n=2$ population leads to an easily observable enhancement in $<\Gamma>$.

Since the escape rate depends on the current through the junction, we carefully measure the escape rate versus time during a pulse and then use Fig. 2 to map from escape rate versus time to current versus time for the pulse. Figure 4 shows results for pulses with successively larger amplitude, from 0.4 V to 0.64 V (the peak voltage of the pulse at the top of the refrigerator was used to characterize the amplitude). To improve statistics, the two smallest pulses were calibrated at high bias currents, $17.81 \mu A$, while larger pulses were calibrated at lower bias currents, $17.74 \mu A$ such that the escape rate during the pulse was always easily experimentally measurable. From Fig. 4, we see that our pulses were quite short, only 1-2 ns in length, and at the peak of the pulse the current is resolvable to 10 nA in 100 ps time intervals.

Figure 5 shows examples of switching curves when the qubit was prepared in different initial states. The open diamonds show the probability $P_e$ that the system escaped versus pulse size for the situation where the qubit was predominantly in its ground state (no resonant microwave excitation applied). For small pulses, the data follow a straight line on a semi-log plot, as expected since in this limit the probability of switching is $P_{01} \sim \Gamma_0 \tau$ and $\Gamma_0$ depends approximately exponentially on current. The solid squares show the corresponding measured switching probability $P_e$ after a resonant microwave pulse of 6.5 GHz was used to produce some population in the first excited state; the measurement pulse was applied 3 ns after the microwaves were turned off to ensure that any undesired transients had decayed away. Finally, the open circles show the corresponding result after a 12.5 GHz resonant microwave pulse was used to produce some population in the second excited state. In this case, $n=1$ was also occupied due to relaxation from $n=2$.

Examination of Fig. 5 reveals clear bending in the switching curves corresponding to the currents at which individual energy levels become highly likely to escape. In order to extract the population in each level, we first use the pulse current calibration (Fig. 4) to determine the current at time $t$, we then find the escape rate at time $t$ from Fig. 2. Starting from an initial occupancy of the levels, we numerically time evolved the occupancies in 0.1 ns increments, accounting for loss due to tunneling from each of the levels. The only variables in this procedure are the initial populations of the levels. The diamonds in Fig. 5 were fit using $P_0 = .99996$, $P_1 = 0$ and $P_2 = 4 \times 10^{-5}$; the squares were fit using $P_0 = 0.80096$, $P_1 = 0.199$ and $P_2 = 4 \times 10^{-5}$; and the circles were fit using $P_0 = 0.867$, $P_1 = 0.088$ and $P_2 = 0.045$. For the 6.5 GHz data (squares), the best chi-square fit for the entire data set came from $P_1 = 0.199$. However, $P_1 = 0.21$ fits the bend more accurately and is plotted as well (not separately visible from $P_1 = 0.199$ in Fig. 5). In general, the agreement is quite good between the fit and the data, with the most serious discrepancies being in the $n=2$ data, probably due to uncertainties in $\Gamma_2$ and the small population in $n=2$.

With these populations known, $P_{10}$ can be found from the measurements. Figure 6 shows the resulting determination of $P_{10}$ (circles), $P_{01}$ (squares), and $N$ (pluses) from the data and fits shown in Fig. 5. The solid circles show $P_{10}$ determined using $P_{01}$ data for parameters $P_0 = 0.80096$, $P_1 = 0.199$. The open circles and plot of $N$ are based on $P_1 = 0.21$, since the behavior near the bend is critical. Examination of Fig. 6 reveals that the optimum single-shot pulse occurred for an amplitude of 135 nA, which yielded an optimum error of $N = 0.03$ and optimal fidelity $F = 94\%$. This error matches the predicted error from Fig. 2 for $u = 100$, which is roughly consistent with the expected ratio of $\Gamma_1/\Gamma_0$ at the peak of the pulse at $I = 17.875 \mu A$.

For comparison, Fig. 6 also shows theoretical results (solid and dashed curves) for $P_{01}$, $P_{10}$ and $N$ for the pulses shown in Fig. 4. The $P_{10}$ curve came from time evolving an initial population $P_1 = 1$ for different pulses and the $P_{01}$ curve from an initial population of $P_0 = 1$. As expected for small pulse amplitudes, $P_{10}$ is large, since the pulse isn't large enough to give the first excited state a large escape rate. For large pulse amplitudes, $n = 1$ tunnels rapidly so that $P_{10}$ is small, however the ground state has a very large escape rate so that $P_{01}$ becomes large. Also shown is a solid curve for the resulting average single-shot error $N$ for $P_1 = 0.5$, $P_0 = 0.5$ and $P_2 = 0$. As noted above, the minimum error does not occur when $P_{10} = P_{01}$. We note that $P_{01}$ agrees very well with the data, while $P_{10}$ shows a small discrepancy, which may come from errors in $\Gamma_1$ or our scaling of a relatively small population, $P_1 = 0.21$, to get $P_{10}$.

We note that other groups have observed diminished fidelity during pulsed readout and suggested that this was caused by coupling to spurious two-level systems during the pulse [15], [31]. Such effects would cause a decrease in $P_{11}$, or equivalently an increase in $P_{10}$. We see no clear evidence for such effects here. Furthermore, fine spectroscopic measurements on our devices show maximum spurious splittings of 5 MHz or less with a density of about 1 per 100 MHz. Based on our pulse rise time, an individual splitting would be crossed in less than one ps, leading to negligible loss of fidelity.

**IV. Conclusions**

We have examined the performance of a fast current pulse technique for reading out the state of a dc SQUID qubit and compared the results to a simple model. The technique is very rapid, allows time-resolved calibration of the current pulse at the qubit, and should ultimately allow a measurement fidelity as high as about 99% and error rate of 0.4%. Although we have not achieved beyond about 94 % fidelity and 3% error rate, use of a somewhat longer pulse should improve the results and possibly enable the application of the technique to error correction.

This work was supported by the National Science Foundation through the QuBIC Program, the National Security Agency, and the state of Maryland through the Center for Superconductivity Research. F.C.W. would like to thank R. Schoelkopf for helpful discussions on fidelity.

Table 1. Calculated results for optimum pulses for representative values of $u = \Gamma_1/\Gamma_0$, the ratio of the escape rate during the measurement current pulse. $\Gamma_1 \tau$ is length of the optimal pulse in units of the mean escape time from $n=1$ level, $P_{01}$ is the probability of making an error in detecting state $n=0$, $P_{10}$ is the probability of making an error in detecting state $n=1$, $N = (P_{10} + P_{01})/2$ is average rate at which an error is made in measuring a state that is equally likely to be $n=0$ or $n=1$, and $F = 1 - P_{10} - P_{01}$ is the measurement fidelity.

| $u = \Gamma_1/\Gamma_0$ | $\Gamma_1 \tau$ | $P_{01}$ | $P_{10}$ | $N$ | $F$ |
|---|---|---|---|---|---|
| 100 | 4.6 | 0.046 | 0.0094 | 0.028 | 0.945 |
| 250 | 5.5 | 0.022 | 0.0035 | 0.013 | 0.974 |
| 500 | 6.2 | 0.013 | 0.0018 | 0.007 | 0.986 |
| 750 | 6.6 | 0.009 | 0.0015 | 0.005 | 0.990 |
| 1000 | 6.9 | 0.007 | 0.0013 | 0.004 | 0.992 |

**Figure Captions**

Fig. 1. Device photograph (a) and schematic (b). Measurement pulses and microwaves pass through a 2fF capacitor to the qubit junction.

Fig. 2. Escape rates of a dc SQUID phase qubit plotted versus bias current. The flux and bias current are ramped such that the current through the qubit junction is equal to the bias current. The dashed line is a fit using a single Josephson junction model with $C = 4.3\,pF$ and $I_c = 17.957\,\mu A$, for the ground state. The theoretical escape rates from the first and second excited levels are plotted as well. Using population estimates from pulsed measurements, a single junction model gives a valid estimate of the escape rates. The large feature in the escape rate at 17.755 µA appears to be due to a very small population in the second excited state ( about 0.01%).

Fig. 3. (a) Single-shot signal-to-noise ratio $S/N$ from Eq. [18] and (b) average error $N$ from Eq. [17] for optimized square adiabatic pulses, plotted as a function of the ratio of the escape rates $\Gamma_1/\Gamma_0$. As expected, the larger the ratio between the escape rates the smaller the detection error.

Fig. 4. Measured current through the qubit plotted versus time for nine pulses with successively larger amplitudes as inferred from switching data.

Fig. 5. Switching probability $P_e$ versus pulse amplitude for three population distributions. Microwaves were applied to enhance populations in the excited levels of the qubit. The diamonds correspond to the situation when no microwaves were applied and the qubit was predominantly in the ground state. The solid squares show the measured fraction that escaped when 6.5 GHz microwaves were applied to enhance $n=1$ state population. The open circles show data where $n=2$ and $n=1$ excited state populations were produced by pumping with microwaves at 12.5 GHz. As the pulses heights are increased the fraction that escapes has distinct features due to the excited levels tunneling at different rates. The solid lines are fits with the only variable being the populations of $n=0$, 1, and 2. The three solid curves show fits to the ground state data (diamonds) with $P_0 = 0.99996$, $P1 = 0$ and $P_2 = 4 \times 10^{-5}$, the first excited state data (squares) with $P_0 = 0.80096$, $P_1 = 0.199, P_2 = 4 \times 10^{-5}$, and then $P_0 = 0.78996$, $P_1 = 0.21$, $P_2 = 4 \times 10^{-5}$ and the second excited state data (circles) with $P_0 = 0.867$, $P_1 = 0.088$, $P_2 = 0.045$.

Fig. 6. Measured error $P_{01}$ (squares) and inferred average error $N$ (pluses) and $P_{10}$ (circles) found from data shown in Fig. 5. Solid and dashed curves show theoretical curves based on the pulse current calibration. Open circles show $P_{10}$ if a population of $P_1 = 0.21$ is inferred from Fig. 5, closed circles for $P_1 = 0.199$. Note offset zero on y-axis.

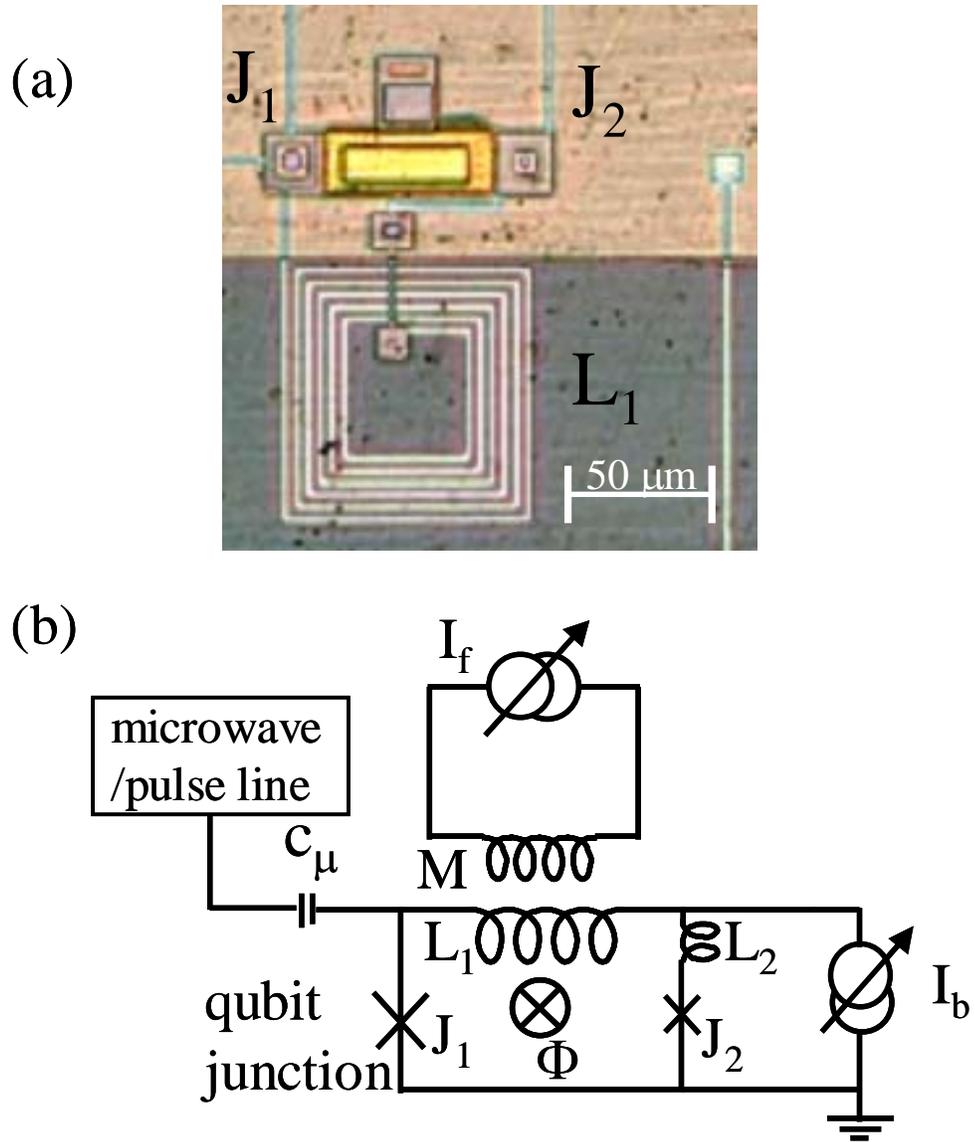

Fig. 1, Palomaki et al.

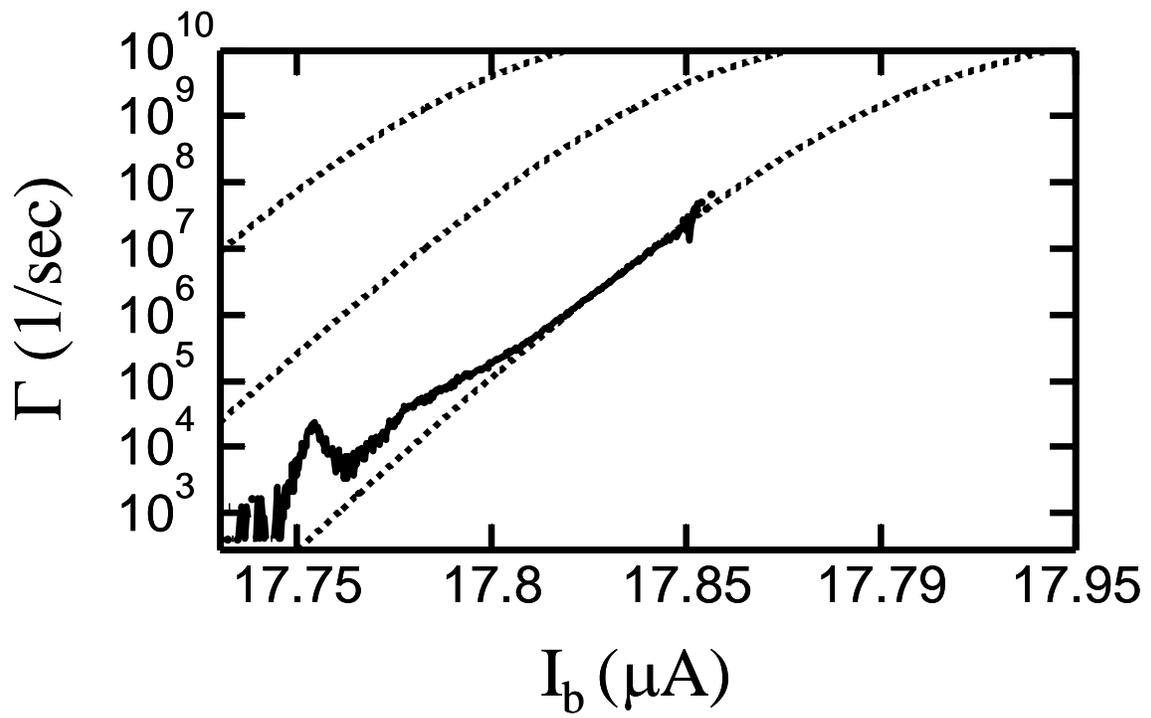

Fig. 2, Palomaki *et al*.

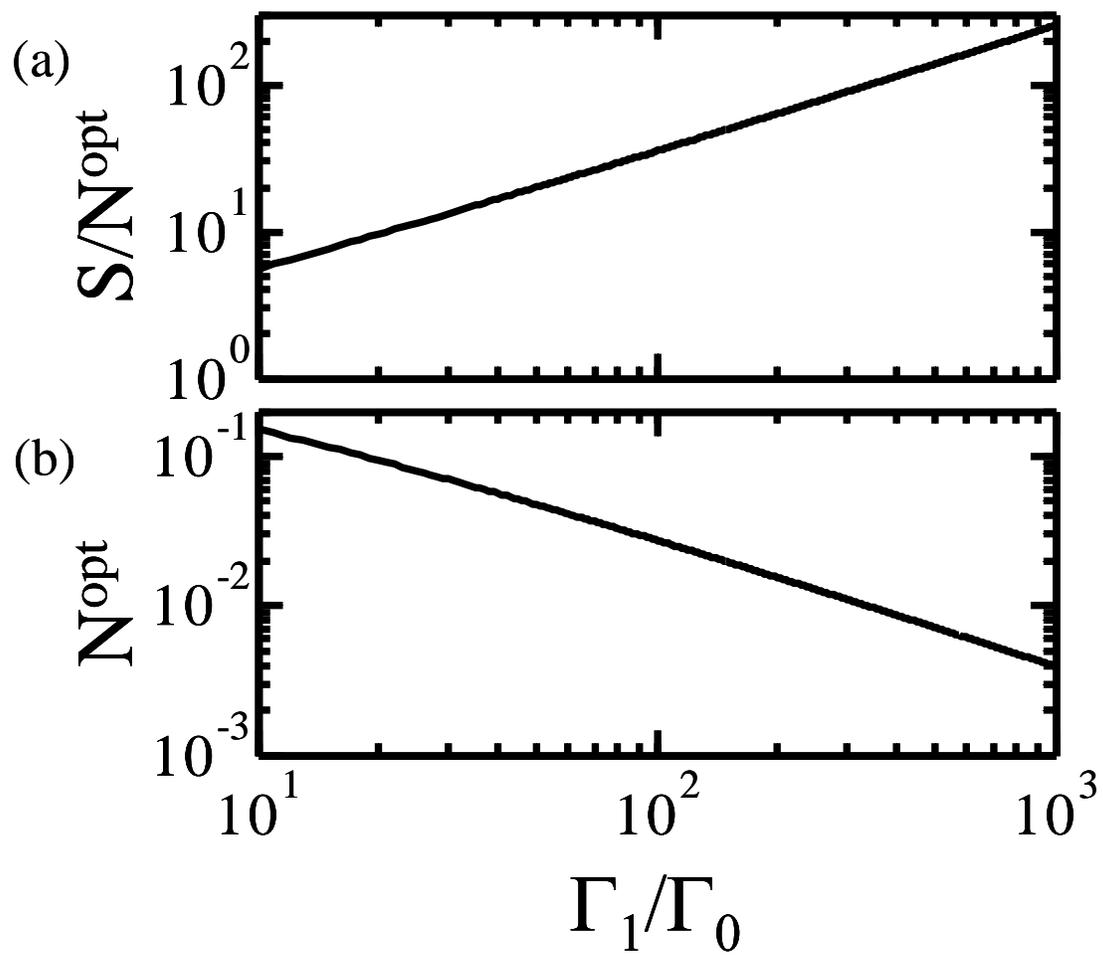

Fig. 3, Palomaki *et al*.

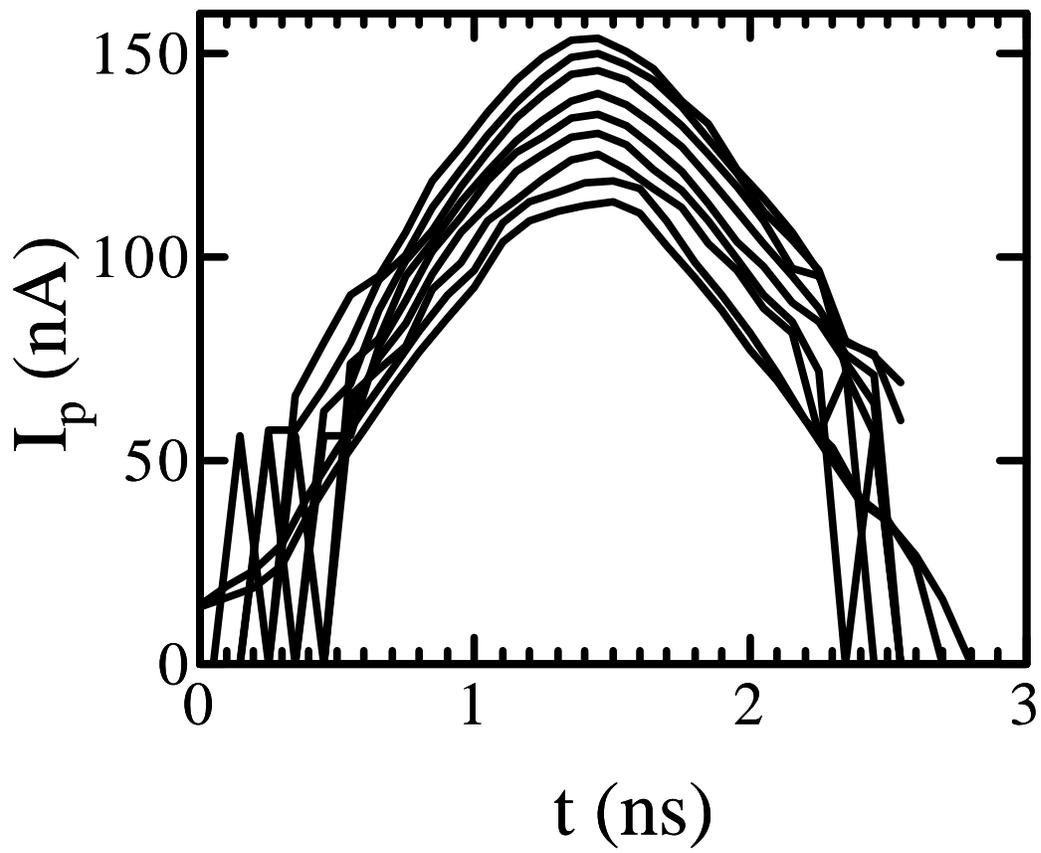

Fig. 4, Palomaki *et al*.

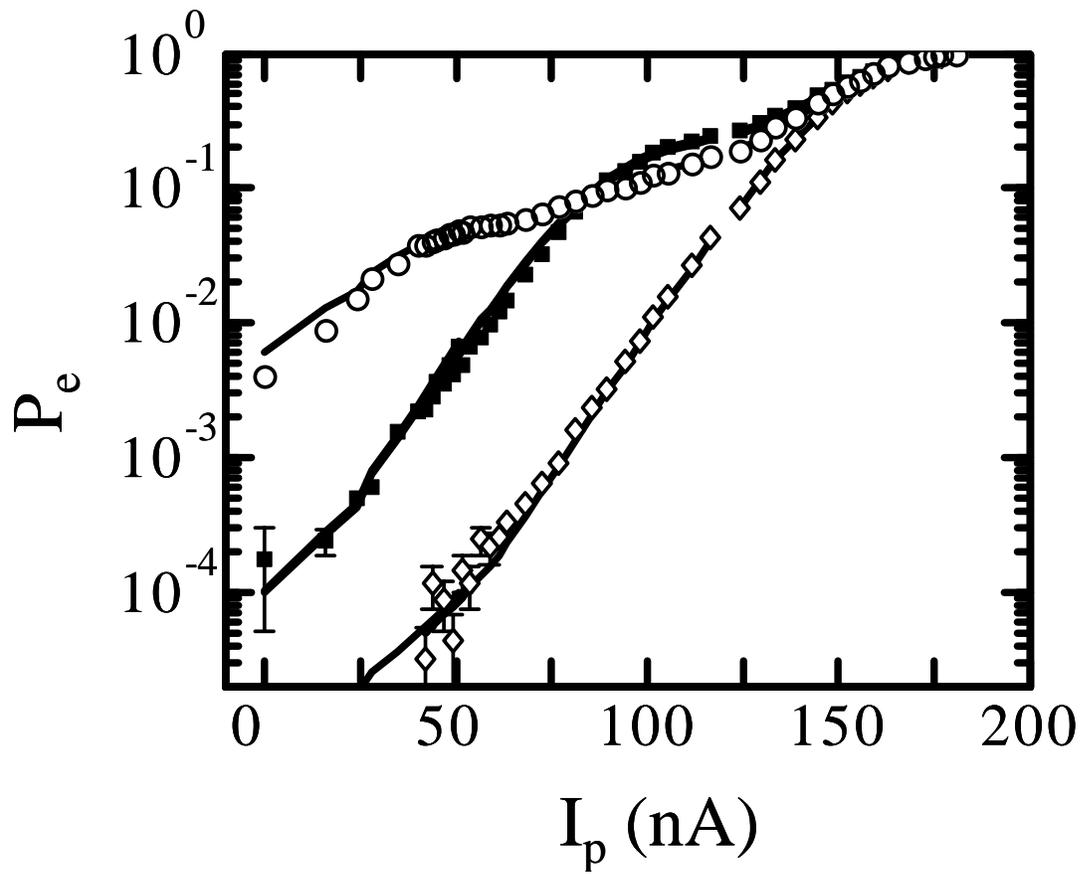

Fig. 5, Palomaki *et al*.

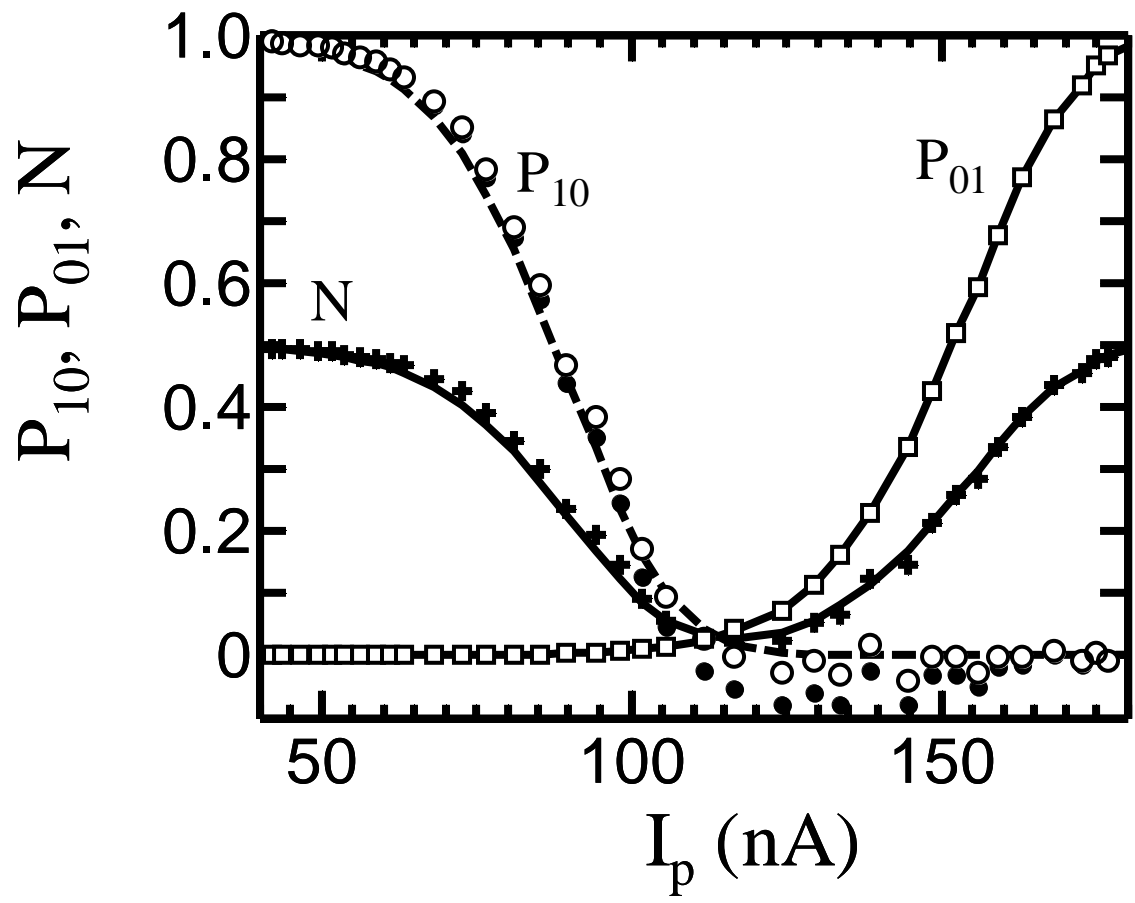

Fig. 6, Palomaki *et al*.